\documentclass[a4paper,aps,prl,twocolumn,groupedaddress]{revtex4}
\usepackage{graphicx}
\usepackage{dcolumn}% Align table columns on decimal point
\usepackage{bm}% bold math

\begin{document}
\title{Dynamics of coupled spins in quantum dots with strong spin-orbit interaction}
\author{A.~Pfund$^{1}$, I.~Shorubalko$^{1}$, K.~Ensslin$^{1}$ and R.~Leturcq$^{1,2}$}

\address{$^1$Solid State Physics Laboratory, ETH Z\"urich, 8093 Z\"urich, Switzerland,
E-mail: pfund@phys.ethz.ch\\
$^2$Institut d'\'Electronique de Micro\'electronique et de Nanotechnologie, CNRS-UMR 8520, Dpt ISEN, Avenue Poincar\'e, BP 60069, 59652  Villeneuve d'Ascq Cedex, France}

\begin{abstract}
We investigated the time dependence of two-electron spin states in a double quantum dot fabricated in an InAs nanowire. In this system, spin-orbit interaction has substantial influence on the spin states of confined electrons. Pumping single electrons through a Pauli spin-blockade configuration allowed to probe the dynamics of the two coupled spins via their influence on the pumped current.
We observed spin-relaxation with a magnetic field dependence different from GaAs dots, which can be explained by spin-orbit interaction. Oscillations were detected for times shorter than the relaxation time, which we attribute to coherent evolution of the spin states.
 \end{abstract}

\maketitle
Double quantum dots (DQDs) are considered as model systems for quantum bits (qubits) in spin-based solid state quantum computation schemes \cite{PhysRevA.57.120}. The combination of single qubit rotations and so-called two-qubit $\sqrt{\mathrm{SWAP}}$ gates would facilitate universal quantum operations.
Fast control of the exchange coupling allows to coherently manipulate coupled spin qubits \cite{PettaScience2005} and to quantify the relevant spin relaxation and coherence times \cite{johnsonPulse,laird:056801} in GaAs based quantum dots.
Beyond the two-qubit operations, controlled rotation of a single spin has been demonstrated \cite{Koppens:2006fk}. 
Especially appealing for a scalable technology is the possibility to perform these single qubit operations with electric gate signals mediated by the spin-orbit interaction (SOI) \cite{NowackScienceEDSR}. This has stimulated the interest in alternative systems with strong spin-orbit interaction, as recently detected in InAs nanowires \cite{fasth:266801,pfund:161308} and carbon nanotubes \cite{KuemmethCNTSOI}.

Complementary to being a tool for single spin rotation, SOI can have substantial influence on two-qubit operations via exchange gates \cite{PhysRevLett.88.047903,PhysRevB.68.115306} or direct spin-spin coupling \cite{trif:085307}. 
Here we investigate the dynamics of two coupled, spatially separated spins in a DQD fabricated in an InAs nanowire, where SOI is orders of magnitudes stronger than in GaAs \cite{fasth:266801,pfund:161308}.
\begin{figure}
\includegraphics[width=0.4\textwidth]{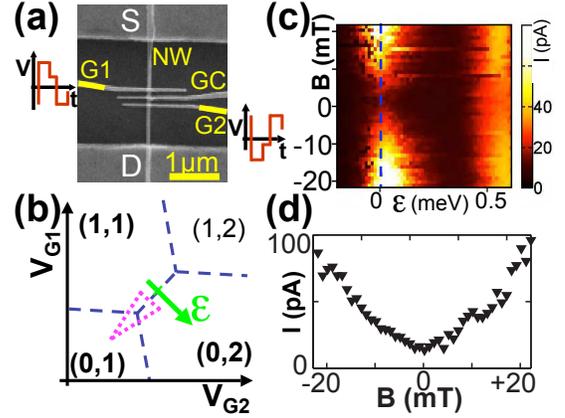}
\caption{(color online). {\bf(a)} Scanning electron microscope image of the measured device. Top-gates $G1$,$G2$ and $GC$ define a double quantum dot in the InAs nanowire (NW) between source $(S)$ and drain $(D)$. Fast voltage pulses can be applied to $G1$ and $G2$. The external magnetic field is parallel to the NW. {\bf(b)} Sketch of a charge stability diagram section of the double dot. Numbers $(n,m)$ label the ground state electron configuration. The axis $\varepsilon$ defines the detuning of the electrochemical potentials in the two dots for 2 electrons in the system. The dotted line indicates the pumping cycle used for the time-dependent measurements.
{\bf(c)} Current $I_{SD}$ for finite bias $V_{SD}=+0.7\,$mV as a function of magnetic field $B$ and detuning $\varepsilon$ at the $(1,1)$-$(0,2)$ transition. Spin-blockade suppresses the current around $B=0$.
{\bf(d)} Cross section along $\varepsilon\approx 0$ as indicated by the dashed line in Fig.\,1(c).}
\end{figure}

We employ a charge pumping scheme \cite{PothierFirstDQDpump,fuhrer:052109} to measure the time dependence of two-electron spin states by transport through the DQD. When the system contains two (excess) electrons, the Pauli exclusion principle suppresses certain transitions \cite{Ono_rectification}. This spin-blockade (SB) can be used to electrically determine the spin state \cite{Koppens:2006fk,johnsonPulse,PettaScience2005,pfund:036801}. The pumped current is strongly reduced in the blockaded direction compared to cycling in the opposite way, which reflects the spin transition rules leading to the SB.
We concentrate on the evolution of those two-electron spin states, where the electrons are distributed between the coupled dots (the $(1,1)$ occupancy). A decay of the SB is observed on a timescale of $\sim\!\!300\,$ns, which we relate to relaxation towards a state with $(1,1)$-triplet character. In contrast, no decay is observed up to several $\mu$s when both electrons occupy the same dot (the $(0,2)$ occupancy). The observed time-dependence differs significantly from measurements in GaAs DQDs and cannot be explained by models accounting only for hyperfine interaction. Instead, the magnetic field dependence is consistent with SOI mediated relaxation \cite{amasha:046803,meunier:126601,Hanson:2005uq}. On a shorter timescale ($\sim\!\!100\,$ns), we detect oscillations between the spin-states. These findings suggest, that coherence times are similar to GaAs DQDs.

We investigate a DQD formed by lithographically defined top-gates on an epitaxially grown InAs nanowire \cite{pfund:161308,pfund:036801}, see Fig.\,1(a). Transport measurements were performed in a dilution refrigerator at an electronic temperature of $130\,$mK. A magnetic field can be applied parallel to the nanowire.
Thermalized coaxial cables allow to apply voltage pulses with a typical rise time of $2\,$ns to the top-gates. Bias tees at low temperature are used to admix AC and DC signals.

The gates $G1,G2$ define tunnel barriers and tune energy levels in dot 1 and 2. The center gate $GC$ separates the two quantum dots. In the presented measurements, the center gate voltage is fixed and defines a tunnel coupling $\approx 3\,\mu$eV. Due to Coulomb blockade, the number of electrons in each dot is fixed for specific regions in the $V_{G1}$-$V_{G2}$-plane \cite{vanderWiel01}. A part of the charge stability diagram is sketched in Fig.\,1(b) and the electronic configuration $(n,m)$ is labeled by the number of electrons n in dot 1 (m in dot 2). These labels refer to the number of excess electrons in addition to spin-less filled shells of electrons \cite{pfund:161308,pfund:036801}. Variation of the gate voltages along the green arrow in Fig.\,1(b) detunes the levels in the dots by an energy $\varepsilon$.

In the case without spin dependent interactions, two electrons form either a singlet $S$ or triplet states $T_{\sigma}$ ($\sigma=0,\pm$ denotes the z-component of the spin state). If the detuning $\varepsilon$ is positive, both electrons are in the same dot and the ground state is the singlet $S(0,2)$. Triplets in $(0,2)$ have higher energies because they involve occupation of an excited orbital state. For $\varepsilon<0$, the singlet $S(1,1)$ and the triplets $T_{0,\pm}(1,1)$ are close in energy at zero magnetic field \cite{Koppens:2005qy}. Since tunneling preserves spin, a transition from a $(1,1)$-triplet to $S(0,2)$ is forbidden. Various experiments show that the singlet-triplet picture describes well the SB in GaAs DQDs \cite{Ono_rectification,johnsonPulse,PettaScience2005,Koppens:2005qy}. In the following, this model is used for a qualitative description.

In Fig.\,1(c) the current through the device is shown as a function of detuning $\varepsilon$ and magnetic field $B$, when no pulses are applied to the gates. A finite source-drain bias $V_{SD}=0.7\,$mV is applied. Sequential transport from $(1,1)$ to $(0,2)$ is in principle allowed, if the relevant levels are within the bias window: $0 \leq \varepsilon \leq |eV_{SD}|$.
Around zero field however, the current in Fig.\,1(c) is strongly suppressed.
In the basic picture described above, blockade arises once a $(1,1)$-triplet is loaded: the state can neither tunnel to $S(0,2)$ nor unload again to the source, if it is within the bias window. 
Not explained by this model is the strong current which sets in for small magnetic fields as shown in Fig.\,1(d). This behavior is not reported in GaAs DQD tuned to the same coupling, but also occurs in other DQDs with strong SOI, as recently found in carbon nanotubes \cite{KuemmethCNTSOI,CNTSOBdependence}. In the following, we identify SOI mediated relaxation to $T_+(1,1)$ as the origin of this difference to GaAs.
\begin{figure}
\includegraphics[width=0.33\textwidth]{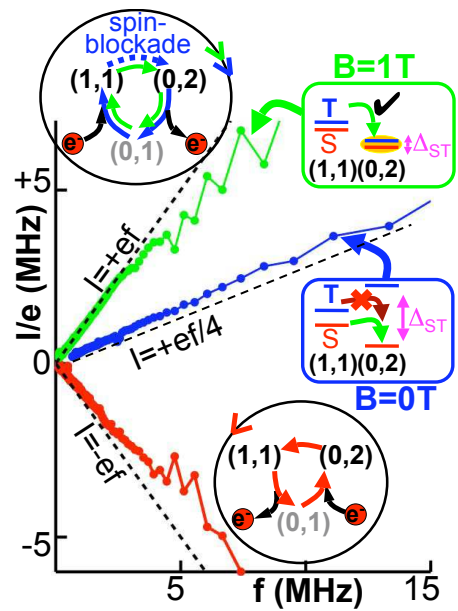}
\caption{(color online). Pumped electrons per time $I/e$ at zero bias as a function of pumping frequency $f$ for cycles as indicated in Fig.\,1(b). The lowest curve (red) shows $I/e$ for anti-clockwise cycling as indicated in the lower round inset. For clockwise cycling (see upper round inset), the pumped current is significantly reduced by Pauli spin-blockade for zero magnetic fields (blue curve) compared to large fields (green curve, $1\,$T). Insets sketch the level energies for the transition $(1,1)$-$(0,2)$ in the clockwise cycle. For B=$0\,$T (blue), spin-blockade suppresses the transition from triplets $T(1,1)$ to the $(0,2)$-singlet. For B=$1\,$T (green), $(0,2)$-singlet and triplet become degenerate and are mixed by spin-orbit interaction. Then no spin-blockade occurs.
 }
\end{figure}

To probe the time evolution of the spin-states, we use pumping cycles where single electrons are shuttled through the DQD \cite{PothierFirstDQDpump,fuhrer:052109}. Fast (ns) pulses are applied to the gates in a loop around the $(0,1)$-$(1,1)$-$(0,2)$-triple point in the charge stability diagram. The voltages are switched rapidly along the dotted line in Fig.\,1(b) and waiting times $t(0,1)$, $t(1,1)$, $t(0,2)$ are spent in each state. The pumped current is measured with zero bias across the device and each point is averaged over $2\,$s.

In Fig.\,2 the pumped current is shown as a function of cycling frequency for cycles with $t(0,1)$=$t(1,1)$=$t(0,2)$. The behavior is different for the two possible pumping directions. The lowest (red) curve shows the result for anti-clockwise cycling (lower round inset). The current is negative and equal to the elementary charge times the cycle frequency up to several MHz as expected. When cycling in the opposite direction (upper round inset), the current is reversed and the pumping efficiency is sensitive to magnetic field. For $B=0\,$T (middle curve, blue), we find a significantly reduced current compared to the anti-clockwise direction. If a high magnetic field $B=1\,$T is applied, charge is again pumped with the full efficiency of one electron per cycle (upper curve, green).

We never observed pumping currents higher than one electron per cycle. 
The tunnel rates in our device correspond to timescales $<1\,$ns (estimated from measurements as in Fig.\,1(c) \cite{pfund:161308}). The pulses are slow with respect to the tunnel rate. Therefore the charge configuration $(n,m)$ during the cycle follows the ground state in the charge stability diagram - provided the transition is not forbidden by spin selection rules.
Beyond that, the pumping efficiency depends on the size of the pulse loop in Fig.\,1(b). 
For example, if the $(0,2)$-corner is chosen at a too high detuning, the transition from $(1,1)$ to $(0,2)$ occurs by electron escape via $(0,1)$ \cite{johnsonPulse}. We adjusted the pulsing parameters so that these processes are minimal.
\begin{figure}
\includegraphics[width=0.45\textwidth]{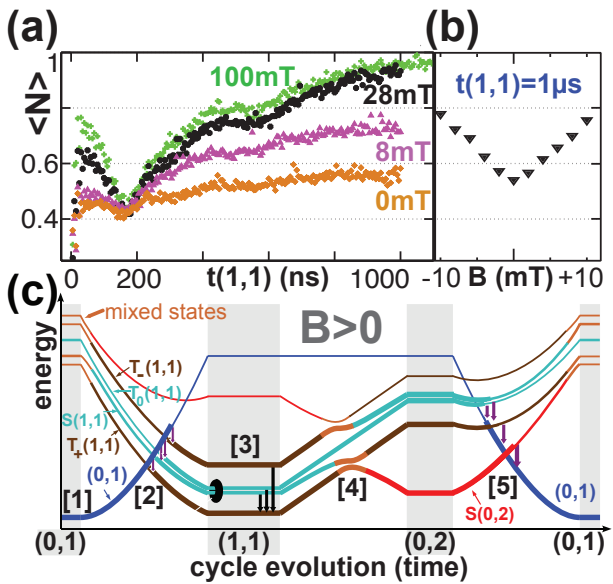}
\caption{(color online). {\bf(a)} Average number of pumped electrons per cycle $\langle N \rangle$ as a function of the time $t(1,1)$ for different magnetic fields. {\bf(b)} Dependence of the long time limit of $\langle N \rangle$ ($t(1,1)=1\,\mu$s) on the external magnetic field $B$. {\bf(c)} Scheme of the energy levels along the pumping cycle for a magnetic field $B > 0$. The system evolves along the thick lines (labels [1]-[5], gray areas represent waiting times): [1] start in $(0,1)$; [2] tunneling of an electron into one of the (1,1) states (arrows); [3] evolution and relaxation in the (1,1) subspace; [4] transition along the detuning axis $\varepsilon$, [5] tunneling out. Only electrons coming from $S(0,2)$ give rise to a pumped current (lowest arrow at [5]). Electrons coming from $(1,1)$-states are only shuttled back and forth (empty arrows). At the transition [4], hyperfine interaction or SOI hybridize different spin states (avoided crossings). During step [3], evolution between the mixed states and relaxation to $T_+(1,1)$ can occur.
}
\end{figure}

The pumping scheme allows to study the time evolution of the quantum states involved in the SB. For a tolerable signal-to-noise ratio of the pumped current, the total cycle times should be shorter than $\approx 2\,\mu$s. Within this limit, we observe no dependence of the pumping efficiency when varying separately the times $t(0,1)$ and $t(0,2)$ (not shown). However, $t(1,1)$ has a strong influence on the pumped current.

In Fig.\,3(a), the average number of pumped electrons per cycle $\langle N\rangle$ is plotted as a function of $t(1,1)$. To maintain a well detectable signal, the total cycle period is fixed to $1.2\,\mu$s. Since $\langle N\rangle$ only depends on $t(1,1)$ for these timescales, we fix $t(0,2)=100\,$ns and compensate the time spent in $(1,1)$ by shortening the time in $(0,1)$ correspondingly. A monotonic long-time increase of $\langle N\rangle$ is found for times $>200\,$ns \footnote{The minimum at $t(1,1)\approx160\,$ns is also observed for different total cycle periods and in schemes, where only $t(1,1)$ is varied. The origin is not fully understood, but it does not affect the analysis of the relaxation process above $200\,$ns.}. At finite field, this effect is much more pronounced than at $B=0\,$T.

The long-time limit of $\langle N\rangle$ is studied as a function of $B$-field in Fig.\,3(b). For $t(1,1)=1\,\mu$s, $\langle N\rangle$ is sensitive to magnetic fields of a few mT. This behavior is in line with the field dependence of the current through SB at finite bias (Fig.\,1(d)).

In order to analyze the behavior of the pumped current, we use the singlet-triplet model for SB \footnote{Since the $(0,2)$ singlet-triplet splitting is much larger then the energy scale of the SOI \cite{pfund:161308}, the $(0,2)$-states are reasonably described as triplets $T_{0,\pm}(0,2)$ and singlet $S(0,2)$. The nature of the $(1,1)$-levels could however be strongly modified by SOI.}.
The values of the pumped currents in Fig.\,2 are related to the spin-transition rules between the corners of the pumping loop.
For the anti-clockwise cycle (lower round inset), the transition from $(0,2)$ to $(1,1)$ is always allowed and one electron is transfered from right to left during each roundtrip.
In the opposite direction (upper round inset), the transition from $(1,1)$ to $(0,2)$ is spin selective. The triplets $T_{0,\pm}(1,1)$ are blocked and only the singlet can pass, which reduces the pumped current.
At $B=1\,$T, the excited triplet $T_+(0,2)$ comes close in energy to the ground state $S(0,2)$ and both are mixed by SOI \cite{pfund:161308}. This way SB is lifted and the full pumping current is recovered.

To understand the decay in Fig.\,3(a), we analyze the spin-selective transition $(1,1)$-$(0,2)$ for different magnetic fields. A contribution to the pumped current is generated only by those $(1,1)$-states, which are transfered into a singlet during the pulse. In other states, the electron is blocked.

At $B=0\,$T, all $(1,1)$-states are close in energy \cite{Koppens:2005qy,PettaScience2005} and become mixed by different spin coupling mechanisms during the time $t(1,1)$. 
The pumped current then reflects the overlap with the singlet. In Fig.\,3(a), the curve for $B=0\,$T shows only a weak time dependence. This supports that there is no preferential evolution towards a certain state, but mixing between all states.

For finite field, the level evolution along the triangular pumping cycle is sketched in Fig.\,3(c). Between the $(1,1)$ and $(0,2)$ corners, triplets and singlet levels would cross at two points (label [4] in Fig.\,3(c)). In the presence of SOI or hyperfine interaction, hybridization of states leads to avoided crossings at these points \cite{Koppens:2005qy,reilly-2008}.

Zeeman splitting lowers the energy of the state with $T_+(1,1)$-character. Relaxation to this new ground state occurs during the time $t(1,1)$. This increases the pumped current, because $T_+(1,1)$ is admixed to the singlet during the charge transition (label [4] in Fig.\,3(c)). We estimate a relaxation time $T_1(1,1) \approx 300\,$ns by fitting with an exponential curve. A comparable relaxation process is not reported in GaAs DQDs, where SB is generally restored with finite magnetic fields \cite{Koppens:2005qy,johnsonPulse,PettaScience2005}.

The B-dependence of $\langle N\rangle$ for long $t(1,1)$ (Fig.\,3(b)) suggests a SOI mediated relaxation. The relaxation rate for these processes generally increases with the splitting of the involved states \cite{amasha:046803,meunier:126601,Hanson:2005uq}. In contrast, spin state decay due to hyperfine interaction with the nuclei is suppressed in a field which splits $T_\pm(1,1)$ \cite{Koppens:2005qy}.
\begin{figure}
\includegraphics[width=0.32\textwidth]{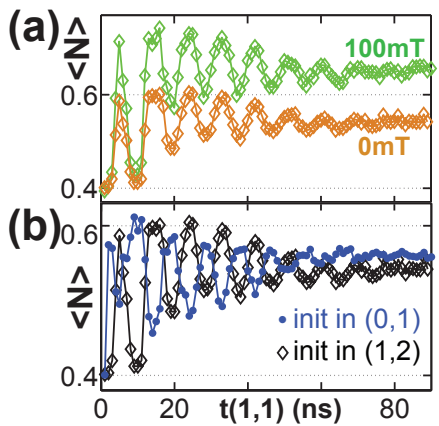}
\caption{(color online). Number of pumped electrons per cycle as a function of $t(1,1)$. {\bf(a)} Dependent on the external field, $\langle N\rangle$ shows oscillations with a period $9.4\,$ns and characteristic decay time of $25\,$ns (for $0\,$mT) and $45\,$ns ($100\,$mT). {\bf(b)} When changing the initialization state of the cycle, the phase of the oscillations changes by $\pi$. Cycles are $(0,1)$-$(1,1)$-$(0,2)$-$(0,1)$ (blue dots) and $(1,2)$-$(1,1)$-$(0,2)$-$(1,2)$ (black rhombs).}
\end{figure}

For times shorter than the relaxation time, the curves in Fig.\,3(a) show up-turns which are not fully understood. However, high resolution measurements in this region reveal striking oscillations of $\langle N\rangle$ as a function of the time $t(1,1)$, as shown in Fig.\,4(a). As above, the total cycle time is constant ($140\,$ns) in a regime, where the signal only depends on $t(1,1)$ ($t(0,2)=20\,$ns fixed).
The oscillation period does not vary with magnetic field, but the decay is changed. A purely exponentially decaying function cannot be fitted to the amplitude. Nevertheless it allows to estimate a decay time, which increases monotonically from $25\,$ns at $0\,$T to $45\,$ns at $100\,$mT.

The oscillations as a function of $t(1,1)$ are robust against variation of the two other waiting times and the total cycle period. The period corresponds to an energy splitting of $h/9.4\,\textrm{ns}=0.44\,\mu$eV, which is consistent with the energy scales for exchange coupling, hyperfine interaction and spin-orbit interaction (at small fields) in the system \cite{pfund:161308}. These energy scales, the magnetic field dependence of the decay and the selective time dependence on $t(1,1)$ suggest coherent evolution in the $(1,1)$ subspace as the origin of the oscillations.

A detection of coherent oscillations in the pumping scheme would imply a selective state preparation. In Fig.\,4(b) we observe a striking dependence of the phase of the oscillations on the way the two-electron state is loaded. Moving the initial state from $(0,1)$ to $(1,2)$ in the charge stability diagram (Fig.\,1(b)) results in a phase shift of $\pi$ (in both cases, the charge is pumped in the direction of SB). These observations suggest that the nature and the coupling of spin-states in DQDs are significantly changed by the SOI compared to the well understood situation in GaAs dots.

By pumping single electrons through a spin-blockaded InAs DQD, we studied the dynamics of two coupled spins in the presence of strong SOI. Beyond the spin-selection rules leading to SB, SOI mediated relaxation to the $(1,1)$-triplet ground state was observed at finite magnetic field. For times shorter than the relaxation time, oscillations were detected in the pumped current. These processes can influence the operation of two-qubit gates in systems with strong SOI.
\begin{acknowledgments}
We thank B.~Altshuler, A.~Imamoglu, D.~Klauser, D.~Loss, C.~Marcus, Y.~Meir, L.~Vandersypen and A.~Yacoby for stimulating discussions, M.~Borgstr\"om and E.~Gini for advice in nanowire growth. We acknowledge financial support from the ETH Zurich.
\end{acknowledgments}
\bibliography{MyBib.bib}

\begin{thebibliography}{26}
\expandafter\ifx\csname natexlab\endcsname\relax\def\natexlab#1{#1}\fi
\expandafter\ifx\csname bibnamefont\endcsname\relax
  \def\bibnamefont#1{#1}\fi
\expandafter\ifx\csname bibfnamefont\endcsname\relax
  \def\bibfnamefont#1{#1}\fi
\expandafter\ifx\csname citenamefont\endcsname\relax
  \def\citenamefont#1{#1}\fi
\expandafter\ifx\csname url\endcsname\relax
  \def\url#1{\texttt{#1}}\fi
\expandafter\ifx\csname urlprefix\endcsname\relax\def\urlprefix{URL }\fi
\providecommand{\bibinfo}[2]{#2}
\providecommand{\eprint}[2][]{\url{#2}}

\bibitem[{\citenamefont{Loss and DiVincenzo}(1998)}]{PhysRevA.57.120}
\bibinfo{author}{\bibfnamefont{D.}~\bibnamefont{Loss}} \bibnamefont{and}
  \bibinfo{author}{\bibfnamefont{D.~P.} \bibnamefont{DiVincenzo}},
  \bibinfo{journal}{Phys. Rev. A} \textbf{\bibinfo{volume}{57}},
  \bibinfo{pages}{120} (\bibinfo{year}{1998}).

\bibitem[{\citenamefont{Petta et~al.}(2005)\citenamefont{Petta, Johnson,
  Taylor, Laird, Yacoby, Lukin, Marcus, Hanson, and
  Gossard}}]{PettaScience2005}
\bibinfo{author}{\bibfnamefont{J.~R.} \bibnamefont{Petta}},
  \bibinfo{author}{\bibfnamefont{A.~C.} \bibnamefont{Johnson}},
  \bibinfo{author}{\bibfnamefont{J.~M.} \bibnamefont{Taylor}},
  \bibinfo{author}{\bibfnamefont{E.~A.} \bibnamefont{Laird}},
  \bibinfo{author}{\bibfnamefont{A.}~\bibnamefont{Yacoby}},
  \bibinfo{author}{\bibfnamefont{M.~D.} \bibnamefont{Lukin}},
  \bibinfo{author}{\bibfnamefont{C.~M.} \bibnamefont{Marcus}},
  \bibinfo{author}{\bibfnamefont{M.~P.} \bibnamefont{Hanson}},
  \bibnamefont{and} \bibinfo{author}{\bibfnamefont{A.~C.}
  \bibnamefont{Gossard}}, \bibinfo{journal}{Science}
  \textbf{\bibinfo{volume}{309}}, \bibinfo{pages}{2180} (\bibinfo{year}{2005}).

\bibitem[{\citenamefont{Johnson et~al.}(2005)\citenamefont{Johnson, Petta,
  Taylor, Yacoby, Lukin, Marcus, Hanson, and Gossard}}]{johnsonPulse}
\bibinfo{author}{\bibfnamefont{A.~C.} \bibnamefont{Johnson}},
  \bibinfo{author}{\bibfnamefont{J.~R.} \bibnamefont{Petta}},
  \bibinfo{author}{\bibfnamefont{J.~M.} \bibnamefont{Taylor}},
  \bibinfo{author}{\bibfnamefont{A.}~\bibnamefont{Yacoby}},
  \bibinfo{author}{\bibfnamefont{M.~D.} \bibnamefont{Lukin}},
  \bibinfo{author}{\bibfnamefont{C.~M.} \bibnamefont{Marcus}},
  \bibinfo{author}{\bibfnamefont{M.~P.} \bibnamefont{Hanson}},
  \bibnamefont{and} \bibinfo{author}{\bibfnamefont{A.~C.}
  \bibnamefont{Gossard}}, \bibinfo{journal}{Nature}
  \textbf{\bibinfo{volume}{435}}, \bibinfo{pages}{925} (\bibinfo{year}{2005}).
  
\bibitem[{\citenamefont{Laird et~al.}(2006)\citenamefont{Laird, Petta, Johnson,
  Marcus, Yacoby, Hanson, and Gossard}}]{laird:056801}
\bibinfo{author}{\bibfnamefont{E.~A.} \bibnamefont{Laird}},
  \bibinfo{author}{\bibfnamefont{J.~R.} \bibnamefont{Petta}},
  \bibinfo{author}{\bibfnamefont{A.~C.} \bibnamefont{Johnson}},
  \bibinfo{author}{\bibfnamefont{C.~M.} \bibnamefont{Marcus}},
  \bibinfo{author}{\bibfnamefont{A.}~\bibnamefont{Yacoby}},
  \bibinfo{author}{\bibfnamefont{M.~P.} \bibnamefont{Hanson}},
  \bibnamefont{and} \bibinfo{author}{\bibfnamefont{A.~C.}
  \bibnamefont{Gossard}}, \bibinfo{journal}{Phys. Rev. Lett.}
  \textbf{\bibinfo{volume}{97}}, \bibinfo{eid}{056801} (\bibinfo{year}{2006}).
  
\bibitem[{\citenamefont{Koppens et~al.}(2006)\citenamefont{Koppens, Buizert,
  Tielrooij, Vink, Nowack, Meunier, Kouwenhoven, and
  Vandersypen}}]{Koppens:2006fk}
\bibinfo{author}{\bibfnamefont{F.~H.~L.} \bibnamefont{Koppens}},
  \bibinfo{author}{\bibfnamefont{C.}~\bibnamefont{Buizert}},
  \bibinfo{author}{\bibfnamefont{K.~J.} \bibnamefont{Tielrooij}},
  \bibinfo{author}{\bibfnamefont{I.~T.} \bibnamefont{Vink}},
  \bibinfo{author}{\bibfnamefont{K.~C.} \bibnamefont{Nowack}},
  \bibinfo{author}{\bibfnamefont{T.}~\bibnamefont{Meunier}},
  \bibinfo{author}{\bibfnamefont{L.~P.} \bibnamefont{Kouwenhoven}},
  \bibnamefont{and} \bibinfo{author}{\bibfnamefont{L.~M.~K.}
  \bibnamefont{Vandersypen}}, \bibinfo{journal}{Nature}
  \textbf{\bibinfo{volume}{442}}, \bibinfo{pages}{766} (\bibinfo{year}{2006}).

\bibitem[{\citenamefont{Nowack et~al.}(2007)\citenamefont{Nowack, Koppens,
  Nazarov, and Vandersypen}}]{NowackScienceEDSR}
\bibinfo{author}{\bibfnamefont{K.~C.} \bibnamefont{Nowack}},
  \bibinfo{author}{\bibfnamefont{F.~H.~L.} \bibnamefont{Koppens}},
  \bibinfo{author}{\bibfnamefont{Y.~V.} \bibnamefont{Nazarov}},
  \bibnamefont{and} \bibinfo{author}{\bibfnamefont{L.~M.~K.}
  \bibnamefont{Vandersypen}}, \bibinfo{journal}{Science}
  \textbf{\bibinfo{volume}{318}}, \bibinfo{pages}{1430} (\bibinfo{year}{2007}).

\bibitem[{\citenamefont{Fasth et~al.}(2007)\citenamefont{Fasth, Fuhrer,
  Samuelson, Golovach, and Loss}}]{fasth:266801}
\bibinfo{author}{\bibfnamefont{C.}~\bibnamefont{Fasth}},
  \bibinfo{author}{\bibfnamefont{A.}~\bibnamefont{Fuhrer}},
  \bibinfo{author}{\bibfnamefont{L.}~\bibnamefont{Samuelson}},
  \bibinfo{author}{\bibfnamefont{V.~N.} \bibnamefont{Golovach}},
  \bibnamefont{and} \bibinfo{author}{\bibfnamefont{D.}~\bibnamefont{Loss}},
  \bibinfo{journal}{Phys. Rev. Lett.} \textbf{\bibinfo{volume}{98}},
  \bibinfo{eid}{266801} (\bibinfo{year}{2007}).

\bibitem[{\citenamefont{Pfund et~al.}(2007{\natexlab{a}})\citenamefont{Pfund,
  Shorubalko, Ensslin, and Leturcq}}]{pfund:161308}
\bibinfo{author}{\bibfnamefont{A.}~\bibnamefont{Pfund}},
  \bibinfo{author}{\bibfnamefont{I.}~\bibnamefont{Shorubalko}},
  \bibinfo{author}{\bibfnamefont{K.}~\bibnamefont{Ensslin}}, \bibnamefont{and}
  \bibinfo{author}{\bibfnamefont{R.}~\bibnamefont{Leturcq}},
  \bibinfo{journal}{Phys. Rev. B} \textbf{\bibinfo{volume}{76}},
  \bibinfo{eid}{161308} (\bibinfo{year}{2007}{\natexlab{a}}).

\bibitem[{\citenamefont{Kuemmeth et~al.}(2008)\citenamefont{Kuemmeth, Ilani,
  Ralph, and McEuen}}]{KuemmethCNTSOI}
\bibinfo{author}{\bibfnamefont{F.}~\bibnamefont{Kuemmeth}},
  \bibinfo{author}{\bibfnamefont{S.}~\bibnamefont{Ilani}},
  \bibinfo{author}{\bibfnamefont{D.~C.} \bibnamefont{Ralph}}, \bibnamefont{and}
  \bibinfo{author}{\bibfnamefont{P.~L.} \bibnamefont{McEuen}},
  \bibinfo{journal}{Nature} \textbf{\bibinfo{volume}{452}},
  \bibinfo{pages}{448} (\bibinfo{year}{2008}).

\bibitem[{\citenamefont{Burkard and Loss}(2002)}]{PhysRevLett.88.047903}
\bibinfo{author}{\bibfnamefont{G.}~\bibnamefont{Burkard}} \bibnamefont{and}
  \bibinfo{author}{\bibfnamefont{D.}~\bibnamefont{Loss}},
  \bibinfo{journal}{Phys. Rev. Lett.} \textbf{\bibinfo{volume}{88}},
  \bibinfo{pages}{047903} (\bibinfo{year}{2002}).

\bibitem[{\citenamefont{Stepanenko et~al.}(2003)\citenamefont{Stepanenko,
  Bonesteel, DiVincenzo, Burkard, and Loss}}]{PhysRevB.68.115306}
\bibinfo{author}{\bibfnamefont{D.}~\bibnamefont{Stepanenko}},
  \bibinfo{author}{\bibfnamefont{N.~E.} \bibnamefont{Bonesteel}},
  \bibinfo{author}{\bibfnamefont{D.~P.} \bibnamefont{DiVincenzo}},
  \bibinfo{author}{\bibfnamefont{G.}~\bibnamefont{Burkard}}, \bibnamefont{and}
  \bibinfo{author}{\bibfnamefont{D.}~\bibnamefont{Loss}},
  \bibinfo{journal}{Phys. Rev. B} \textbf{\bibinfo{volume}{68}},
  \bibinfo{pages}{115306} (\bibinfo{year}{2003}).

\bibitem[{\citenamefont{Trif et~al.}(2007)\citenamefont{Trif, Golovach, and
  Loss}}]{trif:085307}
\bibinfo{author}{\bibfnamefont{M.}~\bibnamefont{Trif}},
  \bibinfo{author}{\bibfnamefont{V.~N.} \bibnamefont{Golovach}},
  \bibnamefont{and} \bibinfo{author}{\bibfnamefont{D.}~\bibnamefont{Loss}},
  \bibinfo{journal}{Phys. Rev. B} \textbf{\bibinfo{volume}{75}},
  \bibinfo{eid}{085307} (\bibinfo{year}{2007}).

\bibitem[{\citenamefont{Pothier et~al.}(1992)\citenamefont{Pothier, Lafarge,
  Urbina, Esteve, and Devoret}}]{PothierFirstDQDpump}
\bibinfo{author}{\bibfnamefont{H.}~\bibnamefont{Pothier}},
  \bibinfo{author}{\bibfnamefont{P.}~\bibnamefont{Lafarge}},
  \bibinfo{author}{\bibfnamefont{C.}~\bibnamefont{Urbina}},
  \bibinfo{author}{\bibfnamefont{D.}~\bibnamefont{Esteve}}, \bibnamefont{and}
  \bibinfo{author}{\bibfnamefont{M.~H.} \bibnamefont{Devoret}},
  \bibinfo{journal}{Europhys. Lett.} \textbf{\bibinfo{volume}{17}},
  \bibinfo{pages}{249} (\bibinfo{year}{1992}).

\bibitem[{\citenamefont{Fuhrer et~al.}(2007)\citenamefont{Fuhrer, Fasth, and
  Samuelson}}]{fuhrer:052109}
\bibinfo{author}{\bibfnamefont{A.}~\bibnamefont{Fuhrer}},
  \bibinfo{author}{\bibfnamefont{C.}~\bibnamefont{Fasth}}, \bibnamefont{and}
  \bibinfo{author}{\bibfnamefont{L.}~\bibnamefont{Samuelson}},
  \bibinfo{journal}{Appl. Phys. Lett.} \textbf{\bibinfo{volume}{91}},
  \bibinfo{eid}{052109} (\bibinfo{year}{2007}).

\bibitem[{\citenamefont{Ono et~al.}(2002)\citenamefont{Ono, Austing, Tokura,
  and Tarucha}}]{Ono_rectification}
\bibinfo{author}{\bibfnamefont{K.}~\bibnamefont{Ono}},
  \bibinfo{author}{\bibfnamefont{D.~G.} \bibnamefont{Austing}},
  \bibinfo{author}{\bibfnamefont{Y.}~\bibnamefont{Tokura}}, \bibnamefont{and}
  \bibinfo{author}{\bibfnamefont{S.}~\bibnamefont{Tarucha}},
  \bibinfo{journal}{Science} \textbf{\bibinfo{volume}{297}},
  \bibinfo{pages}{1313} (\bibinfo{year}{2002}).

\bibitem[{\citenamefont{Pfund et~al.}(2007{\natexlab{b}})\citenamefont{Pfund,
  Shorubalko, Ensslin, and Leturcq}}]{pfund:036801}
\bibinfo{author}{\bibfnamefont{A.}~\bibnamefont{Pfund}},
  \bibinfo{author}{\bibfnamefont{I.}~\bibnamefont{Shorubalko}},
  \bibinfo{author}{\bibfnamefont{K.}~\bibnamefont{Ensslin}}, \bibnamefont{and}
  \bibinfo{author}{\bibfnamefont{R.}~\bibnamefont{Leturcq}},
  \bibinfo{journal}{Phys. Rev. Lett.} \textbf{\bibinfo{volume}{99}},
  \bibinfo{eid}{036801} (\bibinfo{year}{2007}{\natexlab{b}}).

\bibitem[{\citenamefont{Amasha et~al.}(2008)\citenamefont{Amasha, MacLean,
  Radu, Zumbuhl, Kastner, Hanson, and Gossard}}]{amasha:046803}
\bibinfo{author}{\bibfnamefont{S.}~\bibnamefont{Amasha}},
  \bibinfo{author}{\bibfnamefont{K.}~\bibnamefont{MacLean}},
  \bibinfo{author}{\bibfnamefont{I.~P.} \bibnamefont{Radu}},
  \bibinfo{author}{\bibfnamefont{D.~M.} \bibnamefont{Zumbuhl}},
  \bibinfo{author}{\bibfnamefont{M.~A.} \bibnamefont{Kastner}},
  \bibinfo{author}{\bibfnamefont{M.~P.} \bibnamefont{Hanson}},
  \bibnamefont{and} \bibinfo{author}{\bibfnamefont{A.~C.}
  \bibnamefont{Gossard}}, \bibinfo{journal}{Phys. Rev. Lett.}
  \textbf{\bibinfo{volume}{100}}, \bibinfo{eid}{046803} (\bibinfo{year}{2008}).

\bibitem[{\citenamefont{Meunier et~al.}(2007)\citenamefont{Meunier, Vink, van
  Beveren, Tielrooij, Hanson, Koppens, Tranitz, Wegscheider, Kouwenhoven, and
  Vandersypen}}]{meunier:126601}
\bibinfo{author}{\bibfnamefont{T.}~\bibnamefont{Meunier}},
  \bibinfo{author}{\bibfnamefont{I.~T.} \bibnamefont{Vink}},
  \bibinfo{author}{\bibfnamefont{L.~H.~W.} \bibnamefont{van Beveren}},
  \bibinfo{author}{\bibfnamefont{K.-J.} \bibnamefont{Tielrooij}},
  \bibinfo{author}{\bibfnamefont{R.}~\bibnamefont{Hanson}},
  \bibinfo{author}{\bibfnamefont{F.~H.~L.} \bibnamefont{Koppens}},
  \bibinfo{author}{\bibfnamefont{H.~P.} \bibnamefont{Tranitz}},
  \bibinfo{author}{\bibfnamefont{W.}~\bibnamefont{Wegscheider}},
  \bibinfo{author}{\bibfnamefont{L.~P.} \bibnamefont{Kouwenhoven}},
  \bibnamefont{and} \bibinfo{author}{\bibfnamefont{L.~M.~K.}
  \bibnamefont{Vandersypen}}, \bibinfo{journal}{Phys. Rev. Lett.}
  \textbf{\bibinfo{volume}{98}}, \bibinfo{eid}{126601} (\bibinfo{year}{2007}).

\bibitem[{\citenamefont{Hanson et~al.}(2005)\citenamefont{Hanson, van Beveren,
  Vink, Elzerman, Naber, Koppens, Kouwenhoven, and
  Vandersypen}}]{Hanson:2005uq}
\bibinfo{author}{\bibfnamefont{R.}~\bibnamefont{Hanson}},
  \bibinfo{author}{\bibfnamefont{L.~H.~W.} \bibnamefont{van Beveren}},
  \bibinfo{author}{\bibfnamefont{I.~T.} \bibnamefont{Vink}},
  \bibinfo{author}{\bibfnamefont{J.~M.} \bibnamefont{Elzerman}},
  \bibinfo{author}{\bibfnamefont{W.~J.~M.} \bibnamefont{Naber}},
  \bibinfo{author}{\bibfnamefont{F.~H.~L.} \bibnamefont{Koppens}},
  \bibinfo{author}{\bibfnamefont{L.~P.} \bibnamefont{Kouwenhoven}},
  \bibnamefont{and} \bibinfo{author}{\bibfnamefont{L.~M.~K.}
  \bibnamefont{Vandersypen}}, \bibinfo{journal}{Phys. Rev. Lett.}
  \textbf{\bibinfo{volume}{94}}, \bibinfo{pages}{196802}
  (\bibinfo{year}{2005}).

\bibitem[{\citenamefont{van~der Wiel et~al.}(2002)\citenamefont{van~der Wiel,
  Franceschi, Elzerman, Fujisawa, Tarucha, and Kouwenhoven}}]{vanderWiel01}
\bibinfo{author}{\bibfnamefont{W.~G.} \bibnamefont{van~der Wiel}},
  \bibinfo{author}{\bibfnamefont{S.~D.} \bibnamefont{Franceschi}},
  \bibinfo{author}{\bibfnamefont{J.~M.} \bibnamefont{Elzerman}},
  \bibinfo{author}{\bibfnamefont{T.}~\bibnamefont{Fujisawa}},
  \bibinfo{author}{\bibfnamefont{S.}~\bibnamefont{Tarucha}}, \bibnamefont{and}
  \bibinfo{author}{\bibfnamefont{L.~P.} \bibnamefont{Kouwenhoven}},
  \bibinfo{journal}{Rev. Mod. Phys.} \textbf{\bibinfo{volume}{75}},
  \bibinfo{eid}{1} (\bibinfo{year}{2002}).

\bibitem[{\citenamefont{Koppens et~al.}(2005)\citenamefont{Koppens, Folk,
  Elzerman, Hanson, van Beveren, Vink, Tranitz, Wegscheider, Kouwenhoven, and
  Vandersypen}}]{Koppens:2005qy}
\bibinfo{author}{\bibfnamefont{F.~H.~L.} \bibnamefont{Koppens}},
  \bibinfo{author}{\bibfnamefont{J.~A.} \bibnamefont{Folk}},
  \bibinfo{author}{\bibfnamefont{J.~M.} \bibnamefont{Elzerman}},
  \bibinfo{author}{\bibfnamefont{R.}~\bibnamefont{Hanson}},
  \bibinfo{author}{\bibfnamefont{L.~H.~W.} \bibnamefont{van Beveren}},
  \bibinfo{author}{\bibfnamefont{I.~T.} \bibnamefont{Vink}},
  \bibinfo{author}{\bibfnamefont{H.~P.} \bibnamefont{Tranitz}},
  \bibinfo{author}{\bibfnamefont{W.}~\bibnamefont{Wegscheider}},
  \bibinfo{author}{\bibfnamefont{L.~P.} \bibnamefont{Kouwenhoven}},
  \bibnamefont{and} \bibinfo{author}{\bibfnamefont{L.~M.~K.}
  \bibnamefont{Vandersypen}}, \bibinfo{journal}{Science}
  \textbf{\bibinfo{volume}{309}}, \bibinfo{pages}{1346} (\bibinfo{year}{2005}).

\bibitem[{\citenamefont{Churchill et~al.}(2008)\citenamefont{Churchill, Marcos,
  Kuemmeth, Watson, and Marcus}}]{CNTSOBdependence}
\bibinfo{author}{\bibfnamefont{H.~O.~H.} \bibnamefont{Churchill}},
  \bibinfo{author}{\bibfnamefont{D.}~\bibnamefont{Marcos}},
  \bibinfo{author}{\bibfnamefont{F.}~\bibnamefont{Kuemmeth}},
  \bibinfo{author}{\bibfnamefont{S.~K.} \bibnamefont{Watson}},
  \bibnamefont{and} \bibinfo{author}{\bibfnamefont{C.~M.}
  \bibnamefont{Marcus}}, \bibinfo{howpublished}{Contribution to INTNAN8
  conference} (\bibinfo{year}{2008}).

\bibitem[{\citenamefont{Reilly et~al.}(2008)\citenamefont{Reilly, Taylor,
  Petta, Marcus, Hanson, and Gossard}}]{reilly-2008}
\bibinfo{author}{\bibfnamefont{D.~J.} \bibnamefont{Reilly}},
  \bibinfo{author}{\bibfnamefont{J.~M.} \bibnamefont{Taylor}},
  \bibinfo{author}{\bibfnamefont{J.~R.} \bibnamefont{Petta}},
  \bibinfo{author}{\bibfnamefont{C.~M.} \bibnamefont{Marcus}},
  \bibinfo{author}{\bibfnamefont{M.~P.} \bibnamefont{Hanson}},
  \bibnamefont{and} \bibinfo{author}{\bibfnamefont{A.~C.}
  \bibnamefont{Gossard}} (\bibinfo{year}{2008}).

\end{thebibliography}
\end{document}